\documentclass[aps,prd,12pt,preprintnumbers,groupedaddress,floatfix,nofootinbib]{revtex4}

\usepackage{amsmath}
\usepackage{graphicx}
\usepackage{color}

\def\beq{\begin{eqnarray}}
\def\eeq{\end{eqnarray}}
\def\bea{\begin{eqnarray}}
\def\eea{\end{eqnarray}}

\begin{document}

\preprint{DCP-08-04}

\title{{Implications of the discovery of a Higgs triplet on 
electroweak right-handed neutrinos}}

\author{
Alfredo Aranda,$^{1,2}$\footnote{Electronic address:fefo@ucol.mx}, 
J. Hern\'andez-S\'anchez,$^{2,3}$\footnote{Electronic address:jaimeh@ece.buap.mx}
and P.Q. Hung$^{4}$\footnote{Electronic address:pqh@virginia.edu}} 

\affiliation{$^1$Facultad de Ciencias, CUICBAS, Universidad de Colima,
  Bernal D\'{i}az del Castillo 340, Colima, Colima, M\'exico \\
  $^2$Dual C-P Institute of High Energy Physics \\
  $^3$Fac. de Cs. de la Electr\'onica, BUAP, Av. San Claudio y 18 Sur, C. P.
72570, Puebla, Pue., M\'exico \\
  $^4$Dept. of Physics, University of Virginia, 
  382 McCormick Road, P.O. box 400714, 
  Charlottesville, Virginia 22904 - 4714, USA.}

\date{\today}

\begin{abstract}
Electroweak scale active right-handed neutrinos such as those proposed in a recent model
necessitate the enlargement of the SM Higgs sector to include Higgs triplets with
doubly charged scalars. The search for and constraints on such Higgs sector has implications
not only on the nature of the electroweak symmetry breaking but also on the possibility
of testing the seesaw mechanism at colliders such as the LHC and the ILC.
\end{abstract}
\maketitle

\section{Introduction}
\label{intro} 
What is the nature of the spontaneous symmetry breaking (SSB)
of the Standard Model (SM)? Assuming that SSB is due to the vacuum expectation
value (VEV) of some Higgs field, it goes without saying that one of the
most $-$ if not the most $-$ important question that
the Large Hadron Collider (LHC) could help us answer regards to the nature 
of the Higgs mechanism:
Is it just one Higgs doublet as in the minimal SM? or is it a complex
system involving more than one Higgs doublet and perhaps even Higgs
triplets? If one Higgs doublet is sufficient to provide the right kind
of SSB for the minimal SM, why would one need to invoke a more complicated structure?
How well motivated would or should it be?
The possible presence of a Higgs content that includes triplets promises
to yield a rich ``zoo'' of electroweak (EW) scalars to be probed at the LHC and ILC.

Although the possibility of having Higgs triplets that obey
the quintessential electroweak requirement $\rho=1$ has been studied in detail
by~\cite{Chanowitz:1985ug, georgi}, the question that always remained is: Why does one need it?
In the absence of a direct sign on the nature of the Higgs mechanism, it is certainly fair
to contemplate general scenarios as long as they satisfy the electroweak
precision data constraints. However, it would be more appealing if there
were additional motivations for the use of richer structures such as Higgs triplets. 
Recently, a model has been proposed~\cite{Hung:2006ap} in which the right-handed neutrinos that
participate in the seesaw mechanism are {\em active} in the sense that they
are {\em electroweak nonsinglets}. As such, if they are not too heavy, they can be produced
at colliders with electroweak production cross sections and
characteristic signals such as like-sign dileptons. The seesaw mechanism can be 
tested directly at colliders! In fact, the right-handed neutrinos of~\cite{Hung:2006ap} 
are members of SM doublets of mirror leptons and
their Majorana masses are intrinsically linked
to the electroweak scale through a coupling with a Higgs
triplet that develops an electroweak scale VEV. In this model, the sources of the SM SSB
are not only Higgs doublet(s) but include Higgs triplets as well: the nature
of the SM SSB is intimately linked to the nature of neutrino masses and the
possible experimental discovery of the seesaw mechanism at colliders. One
cannot fail but to notice the interesting complementarity of a discovery
of electroweak scale $\nu_R$'s and that of a Higgs triplet.

In this paper we explore the phenomenology of the model in~\cite{Hung:2006ap}.
The full description of the scalar sector involving the triplet fields can be found
in~\cite{Chanowitz:1985ug,Accomando:2006ga,Gunion:1989ci,Han:2007bk},
here we briefly review the extension of the basic model to include electroweak
neutrinos.

In addition to the SM particle content the model of~\cite{Hung:2006ap} contains the
additional fields shown in table~\ref{tab:fields}. There is also an additional global U(1)$_M$ 
symmetry under which \beq \label{u1m} L_R^M, \ e_L^M \rightarrow e^{i\theta_M}L_R^M, \
e_L^M; \ \ \tilde{\chi} \rightarrow e^{-2i\theta_M}\tilde{\chi}, \ \
\phi_S \rightarrow e^{-i\theta_M}\phi_S \ , \eeq and all other
fields are singlets. This global symmetry was invoked in order to
avoid certain terms as indicated below and was explained in detail in~\cite{Hung:2006ap}. 
It turns out, however, that when this model
is embedded into a Pati-Salam-like quark-lepton unification~\cite{Hung2}, 
this global symmetry is no longer needed since
the absence of the aforementioned terms is guaranteed by the
gauge symmetry of the extended model.

\begin{table} 
  \begin{tabular}{|c|c|c|} \hline
    {\rm Additional \ fields} & SU(2)$_W$ & U(1)$_Y$ \\ \hline \hline
    $L_R^M =\left( \nu_R \ \ e_R^M \right)$ & ${\bf 2}$ & $-1$ \\ \hline
    $\tilde{\chi}=\left(\chi^0 \ \ \chi^+ \ \ \chi^{++} \right)^T$ & ${\bf 3}$ & $2$ \\ \hline
    $\xi=\left(\xi^+ \ \ \xi^0 \ \ \xi^{+} \right)^T$ & ${\bf 3}$ & $0$ \\ \hline
    $e_L^M$ & ${\bf 1}$ & $-2$ \\ \hline
    $\phi_S$ & ${\bf 1}$ & $0$ \\ \hline
  \end{tabular}
  \caption{\label{tab:fields} Additional field content of the model with their transformation
    properties under SU(2)$_W$ and U(1)$_Y$.}
\end{table}

We now briefly comment on the virtues of these assignments:
Note that since $\nu_R$ is not an SU(2)$_L$ singlet, it does not
couple to $\bar{L}_L\tilde{\Phi}$. Instead, the Dirac neutrino mass
comes from the term \beq \label{Dmass} {\cal
L}_S=-g_{sl}\bar{L}_L\phi_SL_R^M + h.c. \, \eeq which leads to
$M_{\nu}^D=g_{sl}v_S$, where $\langle\phi_S\rangle=v_S$ and thus the
neutrino Dirac mass is {\it independent} of the EW scale ~\cite{Hung:2006ap}.
Notice that $\phi_S$ is a {\em singlet} Higgs field.

Active right-handed neutrinos must have a mass $> M_Z/2$ in order not to
contribute to the $Z$ width. This is accomplished with the $Y=2$
triplet $\tilde{\chi}$ through the term \beq \label{MRmass}
g_ML_R^{M,T}\sigma_2\tau_2\tilde{\chi}L_R^M \ , \eeq which leads to
\beq \label{MR} M_R=g_M v_M \,, \eeq with $\langle\chi^0\rangle=v_M$ and where
$v_M = O(\Lambda_{EW})$. This allows to have
EW-scale masses for the right-handed neutrinos without having to
fine-tune the Yukawa coupling $g_M$ to be abnormally small ~\cite{Hung:2006ap}.

An important observation is that the U(1)$_M$ symmetry was
introduced ~\cite{Hung:2006ap} in order to forbid the terms
$g_LL_L^T\sigma_2\tau_2\tilde{\chi}L_L$ and
$L_L^T\sigma_2\tau_2 \tilde{\chi}L_R^M$ at tree level. A similar
result is obtained in an extension of that model ~\cite{Hung2} where
the global U(1)$_M$ is not needed for that purpose. 
The main consequence of this is that
the Dirac mass for the neutrinos comes from $v_S$ exclusively and
the Majorana mass, $M_L$, for the left-handed neutrinos arises at the one-loop level
and can be much smaller than $M_R$.

Taking all of this into consideration one obtains the following
Majorana mass matrix: \beq \label{MM} {\cal M}=\left(
\begin{array}{cc}M_L & m_{\nu}^D \\ m_{\nu}^D & M_R \end{array}
\right) \ , \eeq
where, as we have just mentioned above, $M_L \sim \epsilon (m_{\nu}^D)^2/M_R< 10^{-2} (m_{\nu}^D)^2/M_R$. 

We are interested in the scenario where $g_{sl} \sim$~O($g_M$) and
$v_M >> v_S$. In this case, the eigenvalues of ${\cal M}$ become
$-(g_{sl}^2/g_M)(v_S/v_M)v_S(1-\epsilon)$ and $M_R$, where $\epsilon
< 10^{-2}$. Now, since $v_M \sim \Lambda_{EW}$, and using the bound
$m_{\nu}\leq 1$~eV, we have ~\cite{Hung:2006ap} \beq \label{vs} v_S \approx \sqrt{(1
{\rm eV}) \times v_M}\sim {\rm O}(10^{5-6}{\rm eV}) \ . \eeq

So far the $Y=0$ triplet has not played a role since it does not couple to fermions. 
However, it has been introduced in order to ensure $\rho=1$ at tree level ~\cite{Chanowitz:1985ug}.
Note that in principle the parameter $g_{sl}$ is constrained by the neutrino mass spectrum. 
We work under the premise that the smallness of the Dirac mass is a result of $v_S$  
and not from a very small coupling. Thus, if $g_{sl} \sim$ O(1), then $v_S\sim 10^5$~eV.
This amounts to a hierarchy among the scales $v_S/\Lambda_{EW} \sim 10^{-6}$ which is however
not as severe as the usual hierarchies in GUTS. This has been discussed in full detail 
in~\cite{Hung:2006ap,Hung:2007ez}.

\section{Scalar sector}
The kinetic part of the Higgs Lagrangian is 
\begin{eqnarray} \label{Lhiggs}
  {\cal L}_{kin}=\frac{1}{2}Tr[(D_{\mu}\Phi)^{\dagger}(D^{\mu}\Phi)]+
  \frac{1}{2}Tr[(D_{\mu}\chi)^{\dagger}(D^{\mu}\chi)] +
  |\partial_{\mu}\phi_S|^2\ , 
\end{eqnarray} 
where 
\begin{eqnarray} \label{chi}
  \chi=\left(\begin{array}{ccc} \chi^0 & \xi^+ & \chi^{++} \\ \chi^- &
    \xi^0 & \chi^+ \\ \chi^{--} & \xi^- & \chi^{0*} \end{array} \right) \ , 
\end{eqnarray} 
\begin{eqnarray} \label{derivatives}
  D_{\mu}\Phi & = &
  \partial_{\mu}+ig({\bf W \cdot
    \tau}/2)\Phi-ig^{\prime}\Phi B\tau_3/2 \\
  D_{\mu}\chi & = & \partial_{\mu}\chi+ig{\bf W \cdot t}\chi
  -ig^{\prime}\chi B t_3 \ .
\end{eqnarray}

As mentioned above, we work under the premise that
the hierarchy in neutrino masses comes from the VEV of $\phi_S$. This amounts to
$v_S \sim 10^5$~eV and in turn to a negligible mixing between $\phi_S$ and the other
scalars. In what follows we neglect such mixing.

The potential (for $\Phi$ and $\chi$) to be considered
is~\cite{Chanowitz:1985ug} 
\begin{eqnarray} \label{potential} \nonumber
  V(\Phi,\chi)&=&\lambda_1(Tr\Phi^{\dagger}\Phi-v_2^2)^2+
  \lambda_2(Tr\chi^{\dagger}\chi-3v_M^2)^2 \\ \nonumber &+&
  \lambda_3(Tr\Phi^{\dagger}\Phi - v_2^2+Tr\chi^{\dagger}\chi -
  3v_M^2)^2 \\ \nonumber &+& \lambda_4 ( Tr\Phi^{\dagger}\Phi
  Tr\chi^{\dagger}\chi - 2Tr\Phi^{\dagger} T^i\Phi T^j\cdot
  Tr\chi^{\dagger}T^i\chi T^j)
  \\ &+& \lambda_5[3Tr\chi^{\dagger}\chi \chi^{\dagger}\chi
    -(Tr\chi^{\dagger}\chi)^2] \ . 
\end{eqnarray}

Note that this potential is invariant under $\chi \rightarrow
-\chi$. In order for the potential to be positive semidefinite the
following conditions must be imposed: $\lambda_1 + \lambda_2 + 2
\lambda_3 > 0$, $\lambda_1 \lambda_2 + \lambda_1 \lambda_3 +
\lambda_2 \lambda_3 > 0$, $\lambda_4>0$, $\lambda_5 > 0$.
Furthermore the potential is invariant under 
the global symmetry SU(2)$_L \times$ SU(2)$_R$.

When $\chi$ gets a VEV $\langle\chi\rangle=diag(v_M,v_M,v_M)$ it
breaks the global symmetry SU(2)$_L \times$ SU(2)$_R$ down to the
custodial SU(2)$_C$. It was shown in ~\cite{Chanowitz:1985ug, georgi}
that the structure of the VEV is dictated by the proper vacuum alignment.
Now, using $\langle\Phi\rangle=v_2/\sqrt{2}$,
the $W$ and $Z$ masses can be obtained from Eq.~(\ref{Lhiggs}) and
are given by $M_W=gv/2$ and $M_Z=M_W/\cos\theta_W$, with
\beq \label{VEV} v^2=v_2^2+8v_M^2\,, \eeq
with $v \approx 246\,{\rm GeV}$.  This gives rise to $\rho=1$ at tree level.

A convenient parametrization can be made by defining $\cos\theta_H =
c_H \equiv v_2/v$ and thus $\sin\theta_H=s_H\equiv 2\sqrt{2}v_M/v$.
Using these parameters we can see that $\tan\theta_H=t_H$
characterizes the amount of the $W$ mass coming from either the
doublet or the triplet scalars.

One of the important questions that arises in the model of~\cite{Hung:2006ap}
is the relative magnitude of $v_M$ compared with the electroweak scale
$v \sim 246\,{\rm GeV}$. The reason we are interested in this VEV is because
the right-handed neutrino Majorana mass is $M_R= g_M v_M$
as shown in Eq.~(\ref{MR}) and its search through characteristic signals
such as like-sign dilepton events depends crucially on the knowledge
of $M_R$. As we will see below, the constraints coming from the
scalar sector limit the range of allowed values of $\sin\theta_H=s_H\equiv 2\sqrt{2}v_M/v$,
and, consequently, $v_M$. One cannot fail but to see the deep relationship
between the search for the extended Higgs sector and that for the electroweak-scale
active right-handed neutrinos.

We will use the subsidiary fields: 
\begin{eqnarray} \label{components} \nonumber
  \phi^0 & \equiv & \frac{1}{\sqrt{2}}\left( v_2 + \phi^{0r} + i
  \phi^{0i} \right), \ \ \chi^0 \equiv v_M +
  \frac{1}{\sqrt{2}}\left(\chi^{0r} + i \chi^{0i} \right), \\
  \psi^{\pm} & \equiv & \frac{1}{\sqrt{2}}\left( \chi^{\pm} +
  \xi^{\pm} \right), \ \ \zeta^{\pm}  \equiv \frac{1}{\sqrt{2}}\left(
  \chi^{\pm} - \xi^{\pm} \right) 
\end{eqnarray}
for the complex neutral and charged fields, respectively.

The Goldstone bosons are given by 
\begin{eqnarray} \label{goldstone} 
  G_3^{\pm} =c_H\phi^{\pm}+s_H\psi^{\pm}, \ \ G_3^0 =
  i\left(-c_H\phi^{0i}+s_H\chi^{0i}\right) \ . 
\end{eqnarray}

If the potential preserves the SU(2)$_C$ then the fields get arranged 
in the following manner (based on their transformation properties under 
the custodial SU(2)): 
\begin{eqnarray} \label{scalarfields} 
  {\rm five-plet} &\rightarrow& H_5^{\pm
    \pm}, \ H_5^{\pm}, \ H_5^0 \leftrightarrow \ {\rm degenerate} \\
  {\rm three-plet} &\rightarrow& H_3^{\pm}, \ H_3^0 \leftrightarrow \
  {\rm degenerate} \\ 2 - {\rm singlets} &\rightarrow& H_1^0, \
  H_1^{0\prime} \leftrightarrow \ {\rm Only \ these \ can \ mix} \ ,
\end{eqnarray} 
where 
\begin{eqnarray} \label{scalarsrelations} \nonumber 
  &H_5^{++}&=\chi^{++}, \ \ H_5^+ = \zeta^+, \ \ H_3^+=c_H\psi^+-s_H\phi^+ \ , \\
  \nonumber &H_5^0& =
  \frac{1}{\sqrt{6}}\left(2\xi^0-\sqrt{2}\chi^{0r}\right), \ \
  H_3^0=i\left(c_H\chi^{0i}+s_H\phi^{0i}\right), \\ \nonumber
  &H_1^0&=\phi^{0r}, \\
  &H_1^{0\prime}&=\frac{1}{\sqrt{3}}\left(\sqrt{2}\chi^{0r}+\xi^0\right) \ ,
\end{eqnarray} 
with $H_5^{--}=(H_5^{++})^*$, $H_5^-=-(H_5^+)^*$,
$H_3^-=-(H_3^+)^*$, and $H_3^0=-(H_3^0)^*$.
It is also convenient to express the triplet neutral scalar
$\chi^{0}$ in terms of the above states, namely
\begin{eqnarray} 
  \chi^0 \equiv v_M +\frac{1}{\sqrt{3}}H_1^{0\prime}-\frac{1}{\sqrt{6}}
  H_5^0 + \frac{1}{\sqrt{2}\,c_H}H_3^0 \ ,
\end{eqnarray}
where only physical states have been included. Feynman rules for vector boson couplings can be found
in~\cite{Gunion:1989ci}.

One last comment regarding the scalar potential. As discussed in~\cite{Chanowitz:1985ug} the potential
contains an explicit breaking of the U(1)$_M$ symmetry. This renders the model free of NG bosons and the
$\phi_S$ mass is independent of $v_S$. Furthermore there is a would-be-Majoron with a mass
larger than the Z boson mass.

\section{Couplings to matter}
In the search for the Higgs scalars discussed in this work, it
is important to know what those scalars couple to. The couplings
of this extended Higgs sector can be found in ~\cite{georgi}.
Here we are interested in those couplings which are specific to the
model of mirror fermions of ~\cite{Hung:2006ap}. As we shall see below,
they can give rise to very specific signatures such as lepton-number
violating decays.
In this section we obtain the Feynman rules for scalar fermion
couplings including the mirror fermions. 

In the case of SM fermions,
we have the usual Yukawa interactions 
\begin{eqnarray} \label{smyukawas} 
  {\cal L}_Y = -h_{ij}\bar{\Psi}_{Li} \Phi \Psi_{Rj} + h.c. 
\end{eqnarray}

The Feynman rules obtained from this Lagrangian
become~\cite{Gunion:1989ci} 
\begin{eqnarray} \label{feynmanrulesSM}\nonumber
  g_{H_1^0q\bar{q}} & = & -i\frac{m_q \ g}{2 \ m_W \ c_H} \ \ (q=t,b)
  \\ \nonumber g_{H_3^0t\bar{t}} & =
  & i\frac{m_t  \ g  \ s_H}{2  \ m_W \ c_H}\gamma_5, \\
  g_{H_3^0b\bar{b}}& = & -i\frac{m_b  \ g \ s_H}{2 \ m_W \ c_H}\gamma_5, \\
  \nonumber g_{H_3^-t\bar{b}}& = & i \frac{g  \ s_H}{2\sqrt{2} \ m_W \
    c_H}\left(m_t(1+\gamma_5)-m_b(1-\gamma_5)\right) \ , 
\end{eqnarray} 
where third generation notation is used for quarks and similar expressions
apply to leptons.

For mirror fermions we need to consider the terms 
\begin{eqnarray} \label{mirroryukawas1} 
  {\cal L}_{M1}= -g_l^M \bar{L}_R^M \Phi e_L^M +
  h.c. 
\end{eqnarray} 
and 
\begin{eqnarray} \label{mirroryukawas2} 
  {\cal L}_{M2} = - g_ML_R^{M,T}\sigma_2\tau_2\tilde{\chi}L_R^M \,.
\end{eqnarray}

This leads to the following Feynman rules: from Eq.~(\ref{mirroryukawas1})
one obtains
\begin{eqnarray} \label{feynmanrulesmirrorD} \nonumber 
  &g_{H_3^+\nu_{l}\bar{l}^M}&
  =  i\frac{m_{l}^M \ g \
    s_H}{2\sqrt{2} \ m_W \ c_H}(1-\gamma_5) , \\
  &g_{H_1^0l^M\bar{l}^M} & =  -i\frac{m_{l}^M \ g}{2\sqrt{2} \ m_W
    c_H},
  \\ \nonumber &g_{H_3^0l^M\bar{l}^M} &= i\frac{m_{l}^M \ g \
    s_H}{2\sqrt{2} \ m_W \ c_H} \gamma_5, 
\end{eqnarray}
where 
\begin{eqnarray} \label{mirrortaumass} 
  m_{l}^M = g_{l}^M \frac{v_2}{\sqrt{2}} =
  \frac{\sqrt{2} \ m_w \ c_H \ g_{l}^M}{g} , 
\end{eqnarray}
and from Eq.~(\ref{mirroryukawas2}) we get
\beq \label{feynmanrulesmirrorM} \nonumber 
&g_{H_1^{0\prime}\nu_{R}\nu_R}&
= i\frac{g_M \sigma_2 \otimes (1+\gamma_5)}{2\sqrt{3}} , \\
\nonumber &g_{H_5^{0}\nu_{R}\nu_{R}}&
= -i\frac{g_M \sigma_2 \otimes (1+\gamma_5)}{\sqrt{6}} , \\
&g_{H_3^{0}\nu_{R}\nu_{R}}&
= i\frac{g_M \sigma_2 \otimes (1+\gamma_5)}{\sqrt{2}\,c_H} ,\\
 \nonumber &g_{H_5^+\nu_{R}e^{M,+}}&
= i\frac{g_M \sigma_2 \otimes (1+\gamma_5)}{\sqrt{2}} , \\
 \nonumber &g_{H_3^+\nu_{R}e^{M,+}}&
= i\frac{g_M \sigma_2 \otimes (1+\gamma_5)}{\sqrt{2}\, c_H}. \eeq

There are also couplings of SM leptons with their mirrors through
the term in Eq.~(\ref{Dmass}), i.e. \beq \label{feynmanrulesmixing}
\nonumber & g_{\nu_l\bar{\nu}_l\phi_S^r} & =
-i\frac{g_{sl}}{\sqrt{2}}, \ \ g_{\nu_l\bar{\nu}_l\phi_S^i} =
\frac{g_{sl}}{\sqrt{2}} \gamma_5, \\
& g_{l\bar{l}^M\phi_S^r} & = -i\frac{g_{sl}}{2\sqrt{2}}(1-\gamma_5), \ \
g_{l\bar{l}^M\phi_S^i} = \frac{g_{sl}}{2\sqrt{2}}(1- \gamma_5), \eeq where
we have used the definition $\phi_S = v_S
+\frac{1}{\sqrt{2}}(\phi_S^r + i\phi_S^i)$.

A detailed and complete study of the lepton sector of this model has been
presented in~\cite{Hung:2007ez}. In this paper we concentrate on the
scalar sector phenomenology specific to this model.

\section{Numerical Analysis}

\subsection{Scalar sector}

In this section we explore the parameter space of the model. We begin by studying the 
scalar mass spectrum.

The first observation is that the value of $\sin\theta_H$ has an upper bound coming from
from the constraint~\cite{Haber} $\tan\theta_H \leq 2$. It also has a lower bound coming from
the right-handed neutrino mass scale, i.e. $M_R > m_Z/2$. Since $M_R = g_M\,v_M$, this
translates into $g_M\,v_M < 45.6\,{\rm GeV}$. The lower bound on $\sin\theta_H$ comes from
finding the lowest allowable value for $v_M$. If one uses the simple-minded perturbative
requirement $g_M^2/4\,\pi <1$, one obtains $v_M > 12.9\,{\rm GeV}$.
Thus we restrict our study to the range
\beq
\label{sintheta}
0.15 \leq \sin\theta_H \leq 0.89 \ .
\eeq
Equivalently, Eq.~(\ref{sintheta}) can be expressed in terms of the bounds on $v_M$ and $v_2$ namely
\beq
\label{vM}
12.9\,{\rm GeV} < v_M < 77.4\,{\rm GeV} \ ,
\eeq
\beq
\label{v2}
243.3\,{\rm GeV} > v_2 > 112.2\,{\rm GeV}  \ .
\eeq
As we have mentioned above, the restrictions on $\sin\theta_H$ and consequently on $v_M$,
have interesting consequences on the mass range of the electroweak-scale active
right-handed neutrinos. 

We now consider the parameters in Eq.~(\ref{potential}) and explore two general possibilities: Either
there is no hierarchy among the parameters and treat them on equal footing, or we assume that all
parameters involving triplet fields (including those which mix triplet and doublet fields) are 
suppressed with respect to those that involve only doublet fields. Furthermore whenever a parameter 
is not suppressed it is assumed to be of order one and by this we mean that the parameter is
arbitrarily chosen to be in the range $(0.5 - 2)$.

Bounds from unitarity~\cite{Aaki2007} are incorporated through the following relations:
\begin{eqnarray}
  \label{unimh3}
  m_{H3} & \leq & 400~{\rm {\rm GeV}} \ , \\ \label{unimh5}
  m_{\chi} & \leq & \sqrt{3} \ m_{H3} \ , \\ \label{unilight}
  m_{light} & \leq & 270~{\rm {\rm GeV}} \ ,
\end{eqnarray}
where $m_{light}$ stands for the lightest scalar state. There is also a bound in the $m_{H_1^0}-m_{H_1^{0'}}$
plane due to unitarity. It amounts to require the heavier of the two to be less than $(700 - 550)\, {\rm GeV}$
when the lighter is in the range of $(0 - 300)\, {\rm GeV}$. 

Lastly we incorporate the $115\, {\rm GeV}$ LEP lower bound on the lightest scalar mass, however we also 
contemplate the possibility described in~\cite{LEP} that the lightest Higgs might have escaped
detection and could be very light indeed. 

\begin{center}
  \begin{figure}[ht]
    \includegraphics[width=10cm]{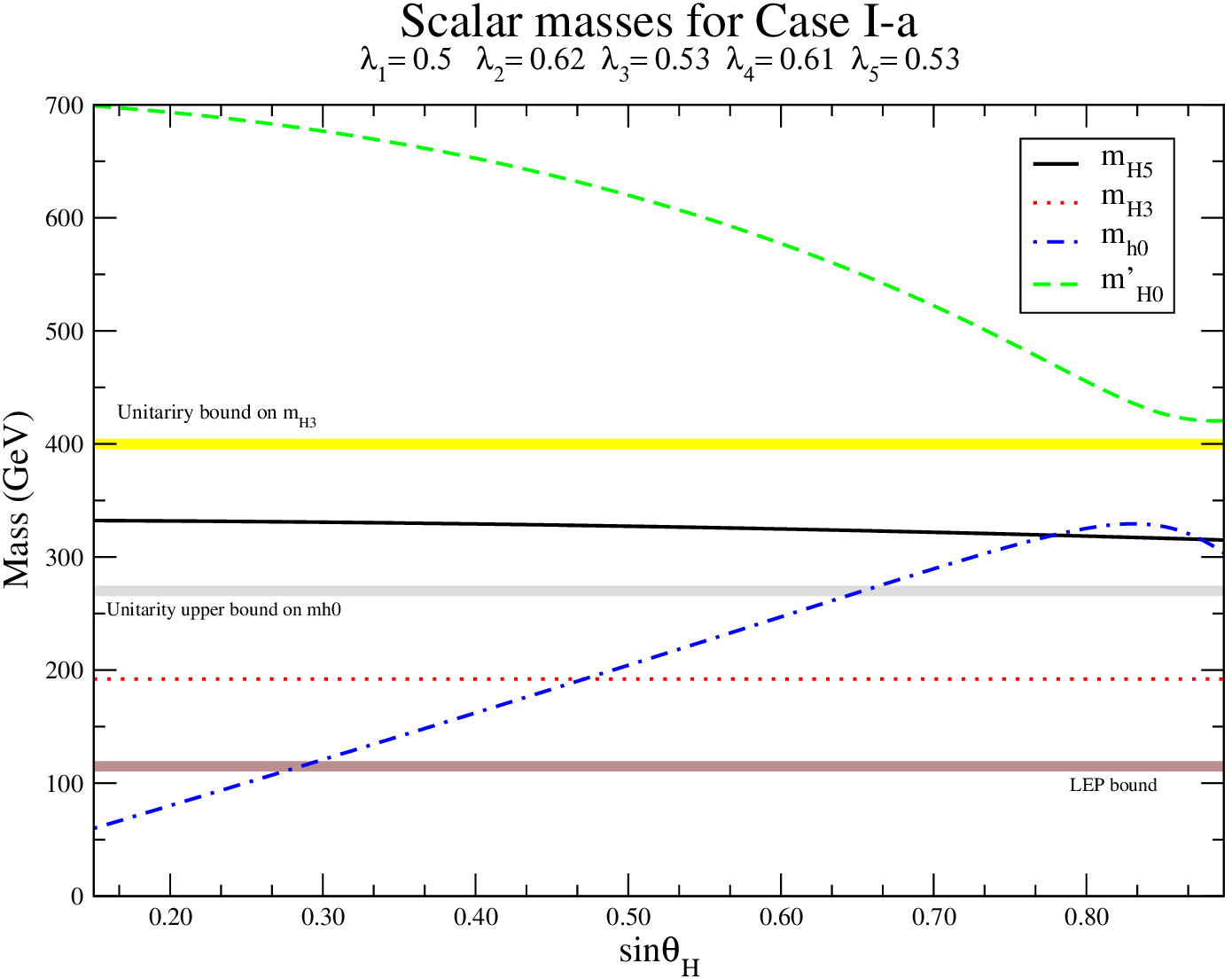}
    \caption{Scalar mass spectrum for the case where there are no hierarchies among
      the parameters in the scalar potential. All parameters are in the lower side of the arbitrarily
      chosen O(1) range. We explicitly show the upper unitarity bound on the lightest scalar
$m_{h^0}$}
    \label{fig:caseIa}
  \end{figure}
\end{center}

We proceed by analyzing some specific cases. Figure~\ref{fig:caseIa} shows the situation when there is no 
hierarchy among the parameters in the scalar potential. They are all of O(1) and taken to be in the lower
part of the arbitrarily chosen O(1) range. It can be seen from the figure that in this case 
the allowed range for $\sin\theta_H$ is $0.3 < \sin\theta_H < 0.65$ where the lower number
refers to the LEP bound while the larger number refers to the unitarity constraint on the mass
of the lightest neutral scalar.
%the complete
%range for $\sin\theta_H$ is allowed.

\begin{center}
  \begin{figure}[ht]
    \includegraphics[width=10cm]{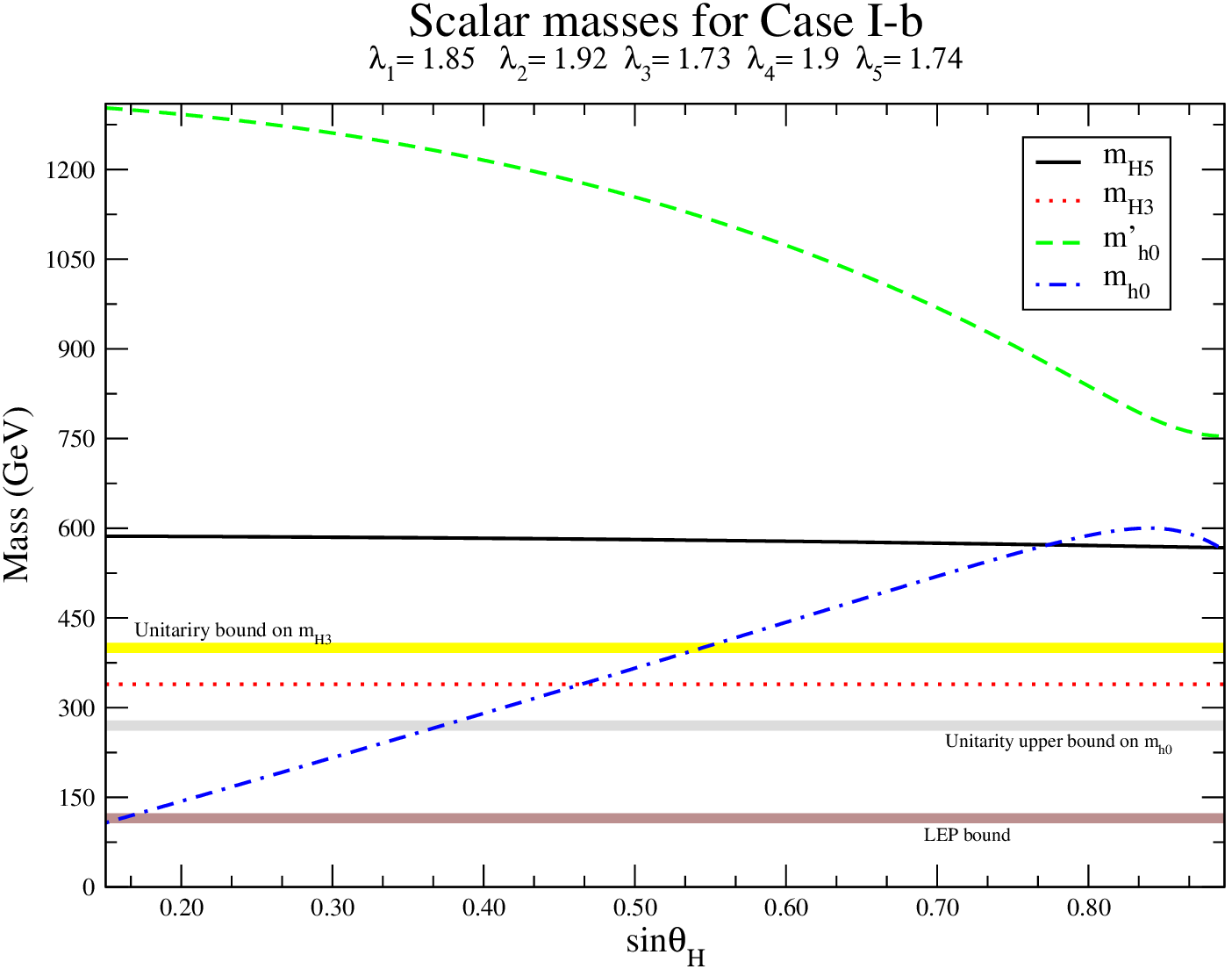}
    \caption{Scalar mass spectrum for the case where there are no hierarchies among
      the parameters in the scalar potential. All parameters are in the upper side of the arbitrarily
      chosen O(1) range.} 
%This case is ruled out by the unitarity bound on $m_{H_1^{0'}}$ discussed in the
%      text.}
    \label{fig:caseIb}
  \end{figure}
\end{center}

Figure~\ref{fig:caseIb} shows a similar case with no hierarchy but with all parameters in the
upper part of the O(1) range. Here the allowed range is shifted downward compared with the
previous bounds, namely $0.17 < \sin\theta_H < 0.35$.

\begin{center}
  \begin{figure}[ht]
    \includegraphics[width=10cm]{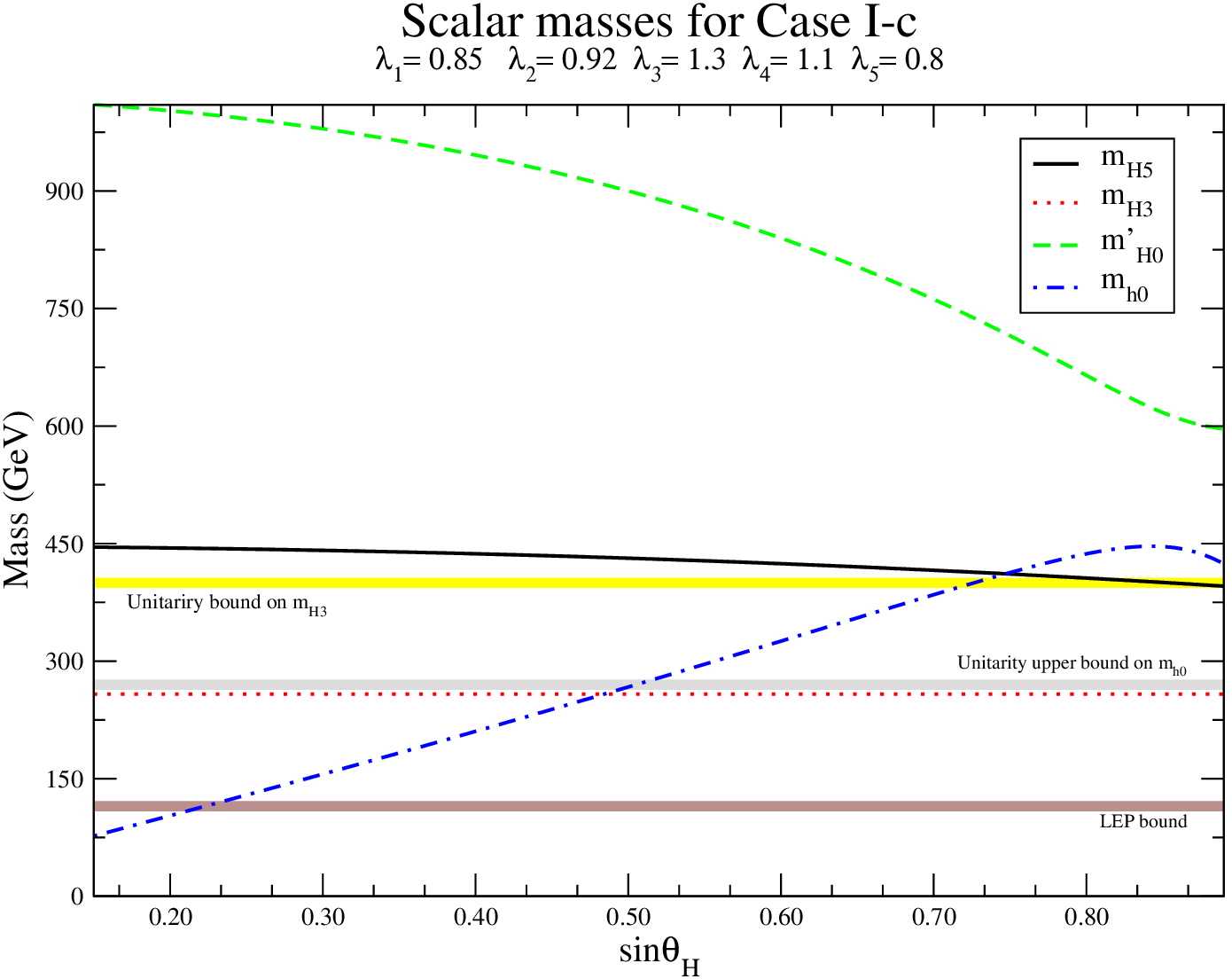}
    \caption{Scalar mass spectrum for the case where there are no hierarchies among
      the parameters in the scalar potential. All parameters have intermediate values in the arbitrarily
      chosen O(1) range.} 
%This case is ruled out by the unitarity bound on $m_{H_1^{0'}}$ discussed in the
%      text.} 
    \label{fig:caseIc}
  \end{figure}
\end{center}

%This scenario is ruled out by the unitarity bound on $m_{H_1^{0'}}$. 
The same situation occurs for the case of intermediate values with no hierarchy as can be seen in 
figure~\ref{fig:caseIc} where now one has $0.22 < \sin\theta_H < 0.48$.
Thus, if all parameters in the potential are taken of the same order, i.e.
no hierarchy, then the allowed range for $\sin\theta_H$ {\em decreases} as those parameters go from
$\sim 0.5$ to $\sim 2$.
%they all have to be in the lower side of the $(0.5- 2)$ range.

\begin{center}
  \begin{figure}[ht]
    \includegraphics[width=10cm]{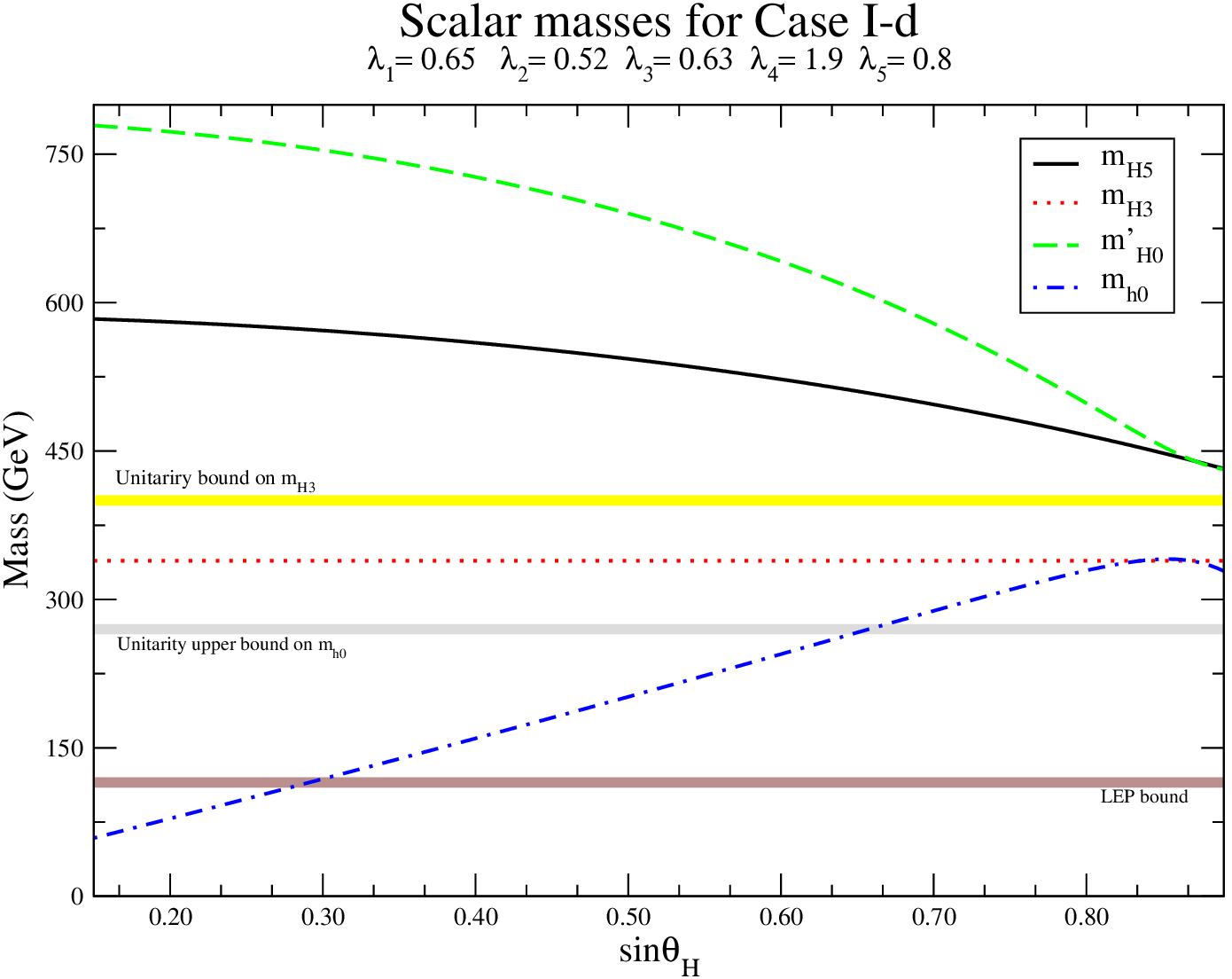}
    \caption{Scalar mass spectrum for the case where there is a small hierarchy between $\lambda_4$
      and the other parameters in the scalar potential. All parameters are in the O(1) range.} 
%This scenario
%      is viable for $\sin\theta_H > 0.7$.}
    \label{fig:caseId}
  \end{figure}
\end{center}

There is an interesting case where we allow for a small hierarchy among some of the parameters, namely
if we let $\lambda_4$ be larger than the other parameters, while still all of them in the O(1) range, then
the situation is that of figure~\ref{fig:caseId}. 
%There we can see that it is possible to satisfy all
%bounds for $\sin\theta_H > 0.7$.

\begin{center}
  \begin{figure}[ht]
    \includegraphics[width=10cm]{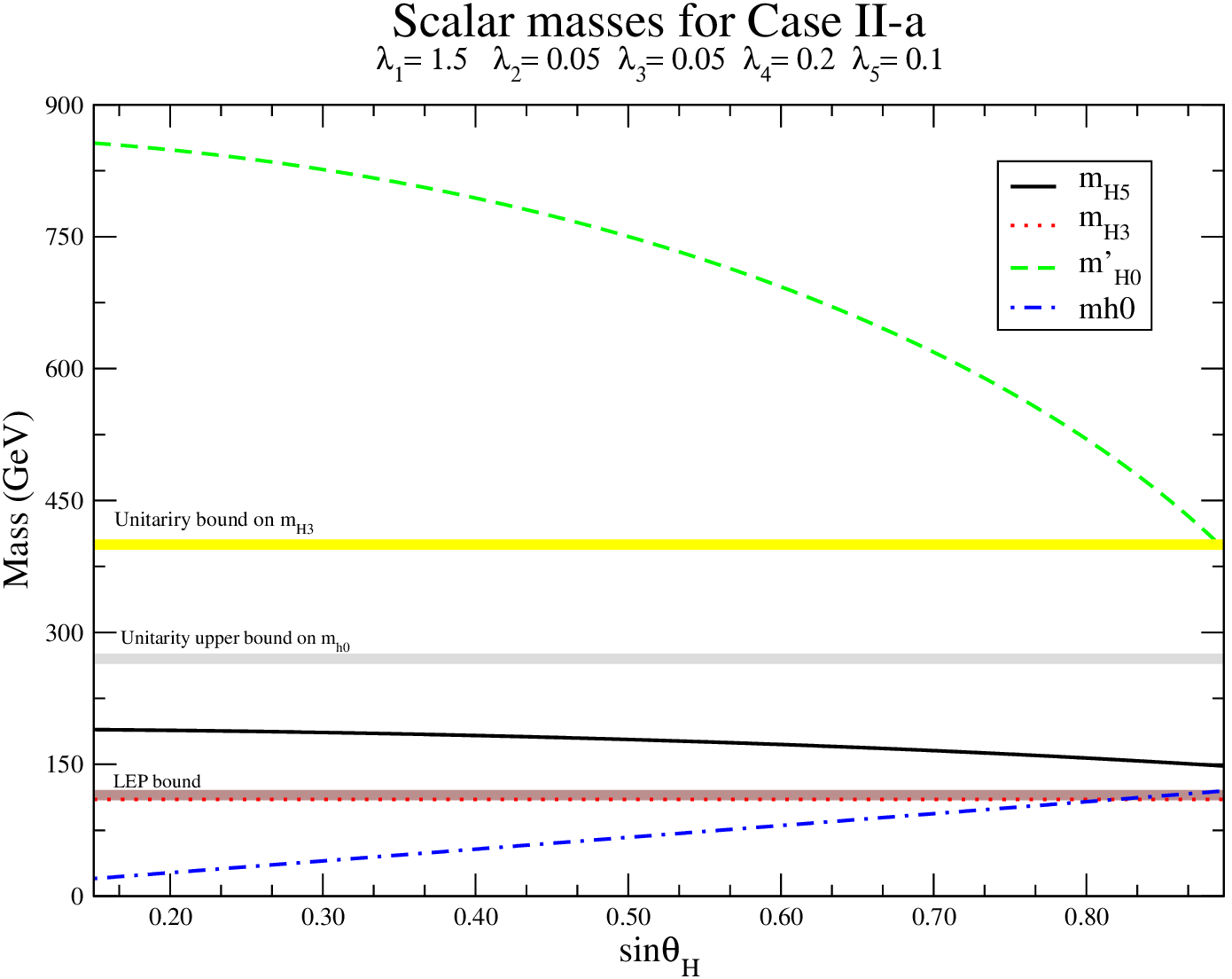}
    \caption{Scalar mass spectrum for the case where there is a hierarchy between $\lambda_1$
      and the other parameters in the scalar potential. $\lambda_1$ lies in the lower part of the
      O(1) range. 
      This scenario satisfies all bounds for $\sin\theta_H > 0.8$,
      except for LEP when $\sin\theta_H < 0.8$ and can be used to analyze the possibility described
      in~\cite{LEP} of an undetected light scalar at LEP.}
    \label{fig:caseIIa}
  \end{figure}
\end{center}

\begin{center}
  \begin{figure}[ht]
    \includegraphics[width=10cm]{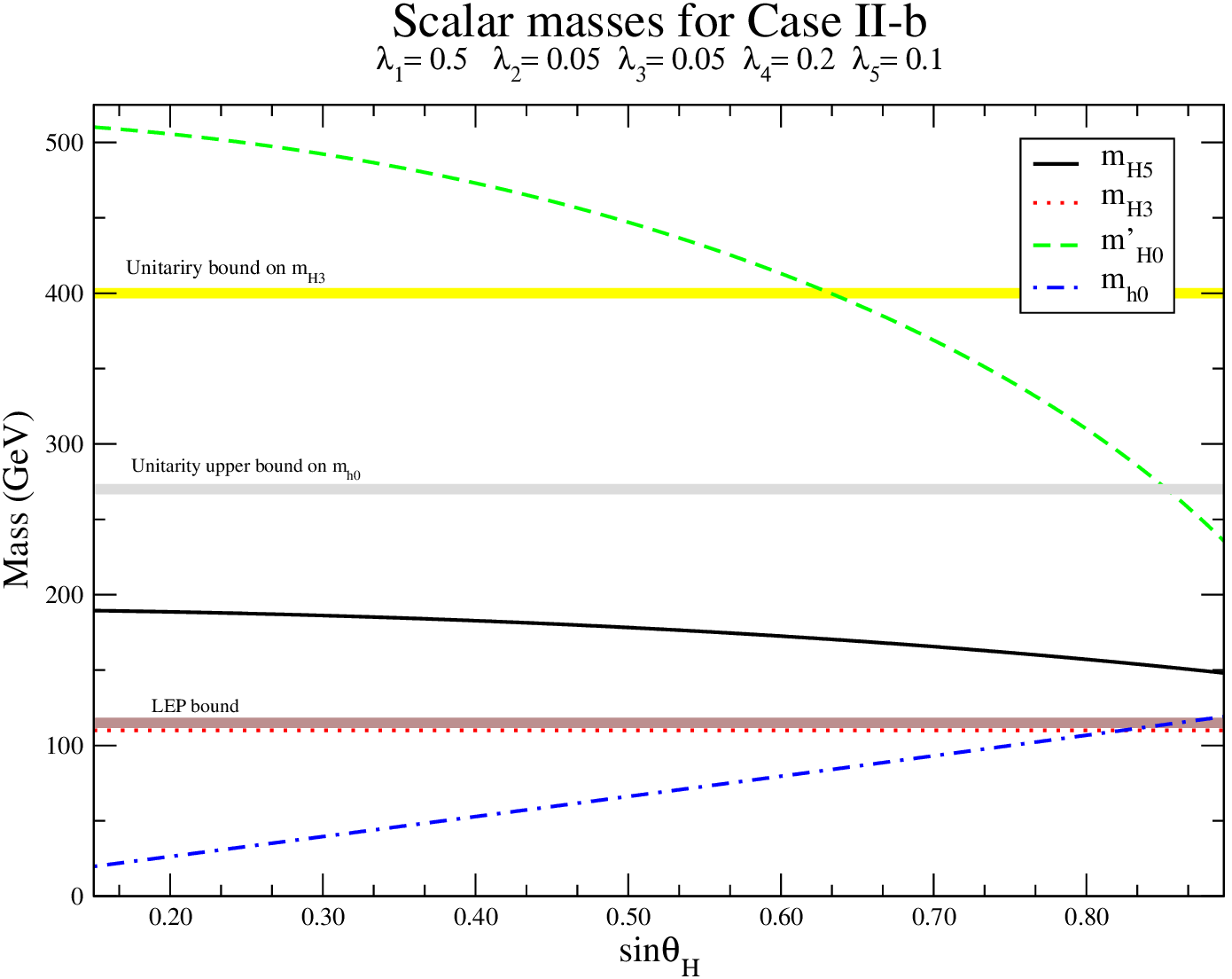}
    \caption{Scalar mass spectrum for the case where there is a hierarchy between $\lambda_1$
      and the other parameters in the scalar potential. $\lambda_1$ lies in the lower range of the 
      O(1) range. 
      This scenario satisfies all bounds
      except for LEP and can be used to analyze the possibility described
      in~\cite{LEP} of an undetected light scalar at LEP.}
    \label{fig:caseIIb}
  \end{figure}
\end{center}

Figures~\ref{fig:caseIIa} and~\ref{fig:caseIIb} show the cases where there is a hierarchy among
$\lambda_1$ and the other parameters. Here $\lambda_1$, which is related to the doublet fields
exclusively, is taken to be of O(1) while the rest are suppressed by a factor of $10$. Again, this factor 
is arbitrary. Figure~\ref{fig:caseIIa} presents the situation where $\lambda_1$ lies in the lower side of the
O(1) range and it can be seen that the spectrum satisfies all bounds for $\sin\theta_H > 0.6$, except for 
the LEP bound. Figure~\ref{fig:caseIIb} shows the result for $\lambda_1$ in the upper part of the O(1) 
range and in this case the spectrum satisfies the bounds (except for LEP) for all the $\sin\theta_H$ range.
It is interesting that these scenarios could fall into the category described in~\cite{LEP} where
there is a light scalar unobserved by LEP. One way to study this possibility is to consider
Higgs production in $e^+ e^-$ collisions, i.e. through the
Higgs-strahlung processes $e^+ e^- \to H_i^0 Z^0$, whose cross sections can be expressed in terms of the
SM Higgs boson (herein denoted by $\phi_{SM}^0$) production formula and the Higgs-$Z^0Z^0$ coupling
as follows~\cite{LEP}:
\begin{eqnarray}
  \sigma_{H_i^0 Z} = R^2_{H_i^0 Z^0 Z^0} \sigma_{H_i^0 Z}^{SM} \ ,
\end{eqnarray}
with
\begin{eqnarray}
  R^2_{H_i^0 Z^0 Z^0}  = \frac{g^2_{H_i^0 Z^0 Z^0}}{g^2_{\phi_{SM}^0 Z^0Z^0}} \ ,
\end{eqnarray}
where $g^2_{H_i^0 Z^0 Z^0}$ is the $H_i^0 Z^0 Z^0 $ coupling in our model
and $g^2_{\phi_{SM}^0 Z^0 Z^0}$ is the $\phi_{SM}^0  Z^0 Z^0$ SM-coupling with the relation
\begin{eqnarray}
\sum_{i=1}^{3} g^2_{H_i^0 Z^0 Z^0} = g^2_{\phi_{SM}^0 Z^0 Z^0} \ .
\end{eqnarray}
In particular, for the lightest scalar in the present model, $R^2_{h^0 Z^0 Z^0}$  is given by:
\begin{eqnarray}
R^2_{h^0 Z^0 Z^0} = \left(-c_H  s_\alpha  + \frac{2 \sqrt{2} }{\sqrt{3}} s_H c_\alpha\right)^2 \ ,
\end{eqnarray}
where $ \alpha $ is the mixing angle that relates the physical states $h^0$, $H^0$ 
to $H_1^0$, $H_1^{'0}$:
\begin{eqnarray}
H_1^0 & = & c_\alpha H^0 -s_\alpha h^0  \ , \\ 
H_1^{'0} & = & s_\alpha H^0 +c_\alpha h^0 \ ,\\
\tan 2 \alpha & = & \frac{2 m^2_{12}}{m^2_{11}-m^2_{12}} \ ,
\end{eqnarray}
where $m_{ij}$ denote the mass-squared matrix elements of the two scalars $H_1^0$, $H_1^{0'}$
given by:
\begin{eqnarray}
 M_{H_1^0,H_1^{0'}}^2 = \left(
\begin{array}{cc}
8 c_H^2 (\lambda_1+\lambda_3) &   2 \sqrt{6}s_{H} c_H \lambda_3  \\
2 \sqrt{6}s_{H} c_H \lambda_3  &  3 s_{H}^2 (\lambda_2+\lambda_3)  \\
\end{array}
\right) \ .
\end{eqnarray}
It is also useful to express the Majorana coupling of $\nu_R$ to the physical
states $H^0$ and $h^0$, namely
\beq \label{feynmanrulesnur} \nonumber 
&g_{H^{0}\nu_{R}\nu_R}&
= i\frac{g_M\,s_\alpha \sigma_2 \otimes (1+\gamma_5)}{2\sqrt{3}} ,\\
 &g_{h^{0}\nu_{R}\nu_R}&
= i\frac{g_M\,c_\alpha \sigma_2 \otimes (1+\gamma_5)}{2\sqrt{3}} .
\eeq

The bounds on the neutral Higgs bosons masses are then expressed in terms of the LEP2 bounds for
$R_{H_i^0 Z^0 Z^0}^2$~\cite{LEP}. We find that large regions of the parameter space of our model are
excluded as can be seen in Table~\ref{tab:1}. By ``parameter space of our model'' we mean the region
in which the mass of the lightest scalar is situated below the LEP bound.
We have defined as ``marginal regions'' those cases that
almost pass the LEP2 bounds on the neutral Higgs mass, i.e.,  when $m_{h^0}
\sim 110\, {\rm GeV}$ and/or when $R_{h^0 Z^0 Z^0}^2$ is almost consistent with the
experimental bounds (see case c in Table~\ref{tab:1}). Our motivation for this definition is that 
once the complete calculation of the one-loop radiative corrections to the mass of the neutral 
Higgs boson is considered, one could expect an enhancement for its mass, thereby allowing it to satisfy
the experimental bounds. It is known that the inclusion of radiative corrections can alter significantly the 
(lightest) neutral CP-even Higgs mass, for example in supersymmetric models as MSSM \cite{Ellis-Ridolfi} and 
MSSM+Higgs triplets\cite{DiazCruz:2007tf}.

%%%%%%%%%%%%%%%%%%%%%%%%%%%%%%%
% Table 2
%%%%%%%%%%%%%%%%%%%%%%%%%%%%%%%%%
\squeezetable
\begin{table*}[htdp]
  \begin{tabular}{|c|c|c|c|c|}
    \hline\hline
    a)  & $  0.34 < s_H < 0.87 $   &
    \begin{tabular}{c}
      45 {\rm GeV} $< m_{h^0} < 116 $ {\rm GeV} 
    \end{tabular} &
    \begin{tabular}{c}
      $0.3 < R_{h^0 Z^0 Z^0}^2 $ for $m_{h^0}<100$ {\rm GeV} \\
      $1.5<R_{h^0 Z^0 Z^0}^2  $ for  $m_{h^0}<116$ {\rm GeV}
    \end{tabular}
    & Excluded by $R_{h^0 Z^0 Z^0}^2 $  \\ \hline
    b) & $ 0.34< s_H < 0.89$  &
    \begin{tabular}{c}
      44 {\rm GeV} $< m_{h^0}<116 $ {\rm GeV}
    \end{tabular}  &
    \begin{tabular}{c}
      $1.5 < R_{h^0 Z^0 Z^0}^2 $ 
    \end{tabular}  &
    \begin{tabular}{c}
      Excluded by $R_{h^0 Z^0 Z^0}^2$
    \end{tabular} \\ \hline
    c) & 0.34 $ < s_H < 0.7$  &
    \begin{tabular}{c}
      55 {\rm GeV} $ < m_{H_1^\pm} <110 $ {\rm GeV}  
    \end{tabular}  &
    \begin{tabular}{c}
      $ 0.19< R_{h^0 Z^0 Z^0}^2 $  for $m_{h^0}<110$ {\rm GeV}\\
      $ R_{h^0 Z^0 Z^0}^2 < 0.29 $ for $m_{h^0} \sim 110$ {\rm GeV}
    \end{tabular}  &
    \begin{tabular}{c}
      Allowed by $R_{h^0 Z^0 Z^0}^2$   \\
      only when $m_{h^0} \sim 110$ {\rm GeV}
    \end{tabular}
    \\  \hline\hline
  \end{tabular}
  \caption{\label{tab:1} Analysis of  $R^2_{h^0 Z^0 Z^0}$ at tree level consistent with
    LEP. We consider experimental limits allowed by LEP2 for charged and
    neutral Higgs bosons for the cases a) $\lambda_1= 1.5$, $\lambda_2=0.05$ and $\lambda_3=0.05$, b)$\lambda_1= 0.5$, 
    $\lambda_2=0.05$ and $\lambda_3=0.05$, c)$\lambda_1= 0.1$, $\lambda_2=0.1$ and $\lambda_3=0.1$.  }
  \label{default1}
\end{table*}
%%%%%%%%%%%%%%%%%%%%%%%%%%%%%%%%

\subsection{Signals from the Higgs triplet neutral scalars}

Interesting and unusual signatures come from the presence of mirror fermions.
In particular, we are interested in signals that show lepton number violation
such as like-sign dilepton events. The Lagrangian in Eq.~(\ref{MRmass}) shows the
coupling of mirror fermions with the $Y=2$ triplet Higgs field. There is no coupling
with the SM leptons which is forbidden either by the U(1)$_M$ symmetry of the model
~\cite{Hung:2006ap} or by embedding it in a Pati-Salam type of quark-lepton unification
~\cite{Hung2}. This coupling which is obviously {\em lepton-number violating} should
show up in the decays of triplet scalars in an interesting way.

One can have the following decays: $H^0,h^0,H^{0}_5, H^{0}_3 \rightarrow \nu_R\,\nu_R$.
The couplings of $\nu_R$ to $H^0,h^0$ are given in Eq.~(\ref{feynmanrulesnur}) and to
$H^{0}_5, H^{0}_3 $ in Eq.~(\ref{feynmanrulesmirrorM}). Depending on the
mass difference between $\nu_R$'s and the charged mirror leptons $e^{M}_R$'s,
the subsequent decay of each $\nu_R$ is $\nu_R \rightarrow e^{M}_R + W^{+} \rightarrow e_L + \phi_S + W^{+}$,
where $e^{M}_R$ could be real or virtual (as well as W's).
Since $\nu_R$ is its own antiparticle one eventually has
$H_{neutral} \rightarrow e^{\mp}_L + e^{\mp}_L +\phi_S + \phi_S + W^{\pm} + W^{\pm}$,
where $H_{neutral} = H^0,h^0,H^{0}_5, H^{0}_3$. This is an example of a
{\em lepton-number violating} like-sign dilepton decay mode of the neutral scalars.
The decay width has a form which is identical to Eq.~(\ref{chitomirrors}) except
for a factor of $1/2$ due to the Majorana nature of $\nu_R$.

\subsection{Signals from $\chi^{++}$ Decays}

The presence of a doubly charged Higgs in this model provides with interesting
phenomenology. Furthermore, the phenomenology of this model is specific and
different from that of the general two triplets model due to the following
observations:

\begin{itemize}
\item Due to the U(1)$_M$ symmetry of the model or its embedding
in a Pati-Salam type of quark-lepton unification, the term proportional to
$l_l^T\sigma_2\tau_2\tilde{\chi}l_L$ is not allowed and thus the decay
$\Gamma(\chi^{++}\to l^+l^+)$ is not present.
\item The presence of mirror fermions and $\phi_S$ allows for the decays $\Gamma(\chi^{++}
\to l_i^M \ l_j^M)$ and $\Gamma(\chi^{++}\to l \ \phi_S \ l_M)$ or
even $\Gamma(\chi^{++}\to l l \phi_S \phi_S)$.
\end{itemize}

We now present the expressions for the relevant $\chi^{++}$ decays.
If $\chi^{++}$ is very heavy, it can have the following decays:
\begin{itemize}
\item $\chi^{++} \to l_M \ l_M$
  \beq \label{chitomirrors}
  \Gamma(\chi^{++} \to l_i^M l_j^M) = \frac{g_M^2 \ m_{\chi}}{16 \pi
    (1 + \delta_{ij})} \left(1 - 4 r_M^2 \right)^{1/2} \ ,
  \eeq
  where $r_M=m_l^M/m_{\chi}$ and then 
  \beq \label{mirror}
  \Gamma(l^M \to l \ \phi_S^r)=\frac{g_{sl}^2 \ m_l^M}{64 \pi} 
  \left(1 - \frac{m_S^2}{(m_l^M)^2} \right)\left|1 - \frac{m_S^2}{(m_l^M)^2} \right| \ .
  \eeq
\item $\chi^{++} \to W^+ W^+$
  \beq \label{chitoww}
  \Gamma(\chi^{++} \to W^+ W^+) = \frac{g^4 v_M^2}{32 \pi r_W^4 m_{\chi}}
  \left(1-4r_W^2 \right)^{1/2} \left(1-4r_W^2+12r_W^4 \right) \ ,
  \eeq
  where $r_W \equiv m_W/m_{\chi}$, and $v_M = \langle \chi^0 \rangle$. 
\item $\chi^{++} \to H_3^+ W^+$
\beq \label{chitoh3w}
\Gamma(\chi^{++} \to H_3^+ W^+) = \frac{c_H^2\,g^2\,m_{H_3}}{32\,\pi\,x^3\,y^2}
F_{1}(x,y)\,F_2(x,y) \,,
\eeq
where $x\equiv m_{\chi}/m_{H_3}$, $y \equiv m_{W}/m_{H_3}$, and
\beq \label{F12} \nonumber
&F_1(x,y)&
= 1 + x^4 -3 y^2 + 2 y^4 - 2 x^2(1+y^2) , \\
\nonumber &F_2(x,y)&
= (x^4 + (y^2 -1)^2 - 2 x^2(1+y^2))^{1/2} .
\eeq
  %\beq \label{chitoh3w}
  %\Gamma(\chi^{++} \to H_3^+ W^+) = \frac{1}{2t_H^2}\Gamma(\chi^{++} \to W^+ W^+).
  %\eeq
\end{itemize}

For intermediate $\chi^{++}$ masses we can have the following three body decays (most
relevant ones):
\begin{itemize}
\item $\chi^{++} \to W^{+*} W^+ \to W^+ \ l^+ \ \nu_l$
  \beq \label{wlnu}
  \Gamma(\chi^{++} \to l^+ \ \nu \ W^+)= \frac{g^4 s_H^2 m_{\chi}}{12 (8\pi)^3\,r_W^4}
  \left(1-4r_W^2+24r_W^4\right) \ .
  \eeq
\item $\chi^{++} \to W^{+*} H_3^+ \to H_3^+ \ l^+ \ \nu_l$
  \beq \label{H3lnu}
  \Gamma(\chi^{++} \to l^+ \ \nu \ H_3^+)= \frac{g^2 c_H^2 m_{\chi}}{12(8\pi)^3\,r_W^4}
  \left( 1-12r_{H3}^2 \right) \ ,
  \eeq
  with $r_{H3}=m_{H3}/m_{\chi}$.
\item $\chi^{++} \to l_M^{*} l_M \to l_M \ l^+ \ \phi_S$ 
  \beq \label{lphilm}
  \Gamma(\chi^{++} \to l^+ \ \phi_S \ l_M)= \frac{3 g_M^2 g_{sl}^2 m_{\chi}}{(16\pi)^3\,r_M^4}
  \left( 1+4r_M^2-3r_S^2 \right) \ ,
  \eeq
  where $r_S=m_S/m_{\chi}$.
\end{itemize}      

\subsubsection{Branching Ratios}

Using the previous expressions we can compute the branching ratios. In the following analysis
we have made the following assumptions:

\begin{itemize}
\item $g_M$ and $g_{sl}$ are proportional to the identity matrix and so, in each of the
expressions above, $g_M$ and $g_{sl}$ represent numbers.
\item The model requires $g_{sl}^2/g_M \sim$~O(1). We have chosen numbers of O(1) for
both couplings and for the numerical results presented below they have been set to
$g_M = 0.7$ and $g_{sl} = 0.8$.
\end{itemize}

Given these assumptions we compute the following branching ratios: 
$B(\chi^{++}\to l^+_M l^+_M)$, $B(\chi^{++}\to W^+ W^+)$, $B(\chi^{++}\to H^+_3 W^+)$,
$B(\chi^{++}\to l^+ \nu W^+)$ and $B(\chi^{++}\to l^+ \phi_S l_M^+)$. Note that
from Eq.~(\ref{H3lnu}) we could compute the corresponding branching ratio, however
in order to satisfy the unitarity condition in Eq.~(\ref{unimh3}) this decay cannot take
place in the model.

\begin{center}
  \begin{figure}[ht]
    \includegraphics[width=5cm]{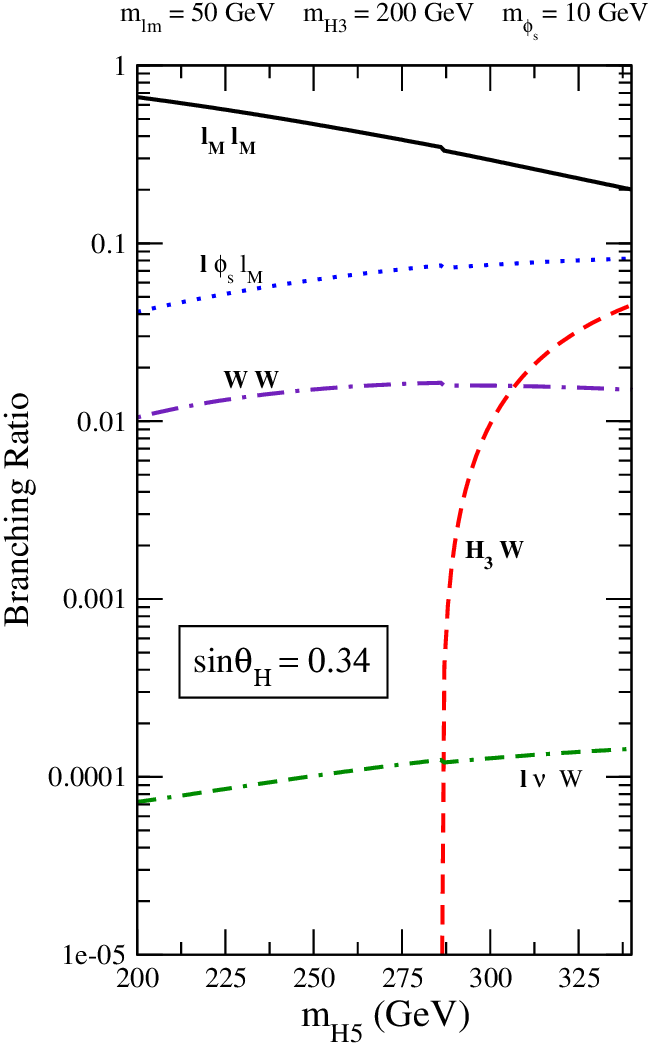}
    \includegraphics[width=5cm]{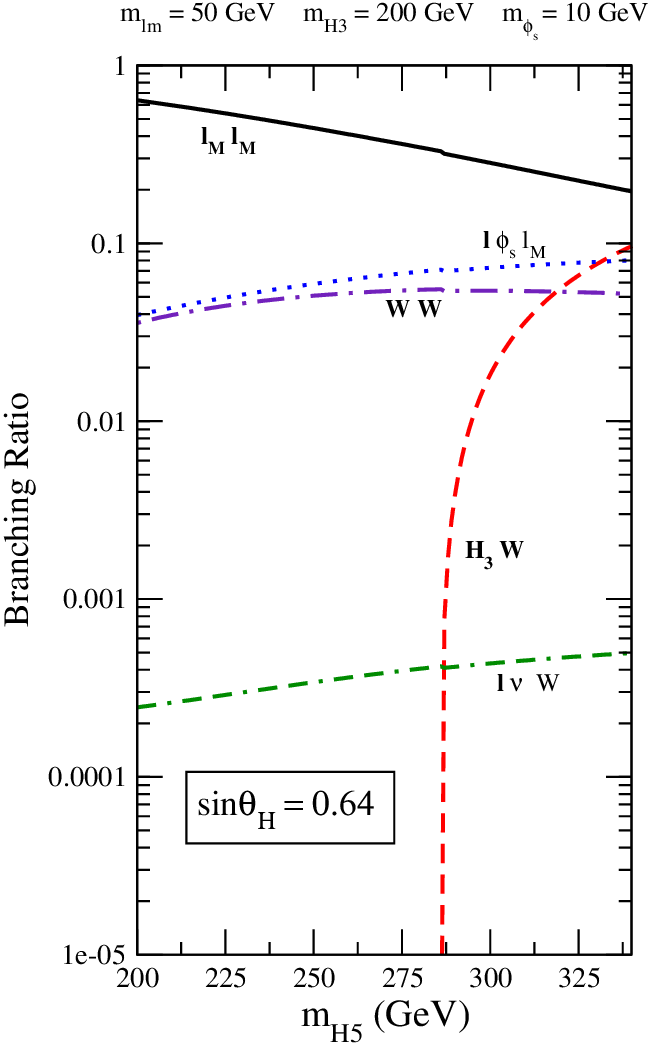}
    \includegraphics[width=5cm]{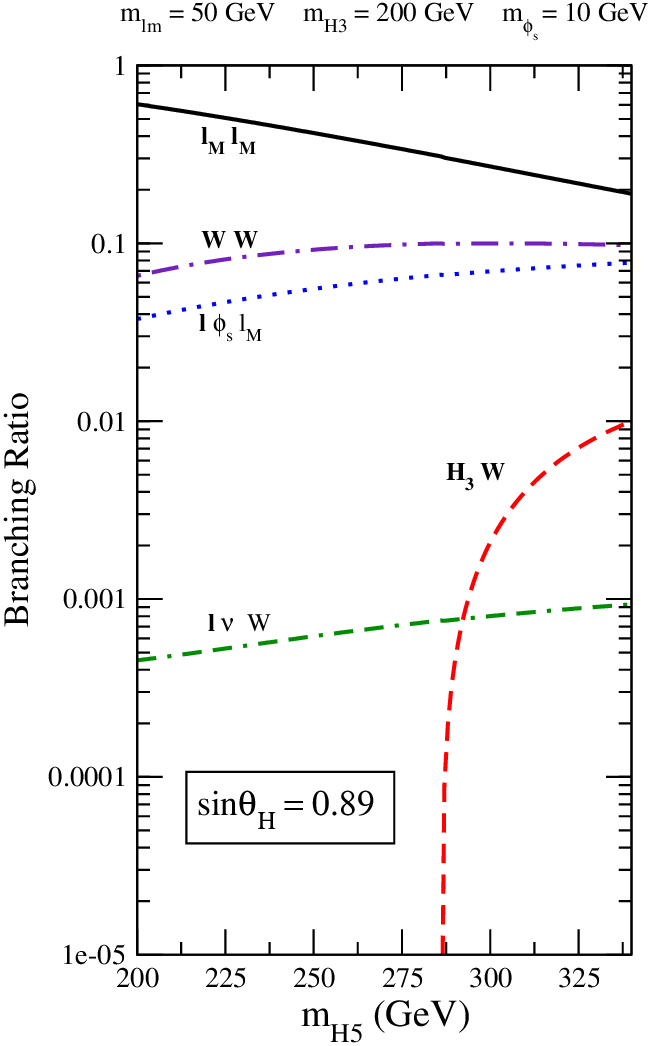}
    \caption{Branching ratios for $\chi^{++}$ as a function of its mass,
      for three different values of $\sin\theta_H$, and for a small $m_{lM}$.}
    \label{fig:br1}
  \end{figure}
\end{center}

Figure~\ref{fig:br1} shows the branching ratios for three different values
of $\sin\theta_H$ and for small values of the mirror fermions masses (taken
to be degenerate) $m_{lM} = 50\, {\rm GeV}$. We can see that the dominant one always
corresponds to $B(\chi^{++}\to l_Ml_M)$, while the relative dominance of
the other channels depends on $\sin\theta_H$.

\begin{center}
  \begin{figure}[ht]
    \includegraphics[width=5cm]{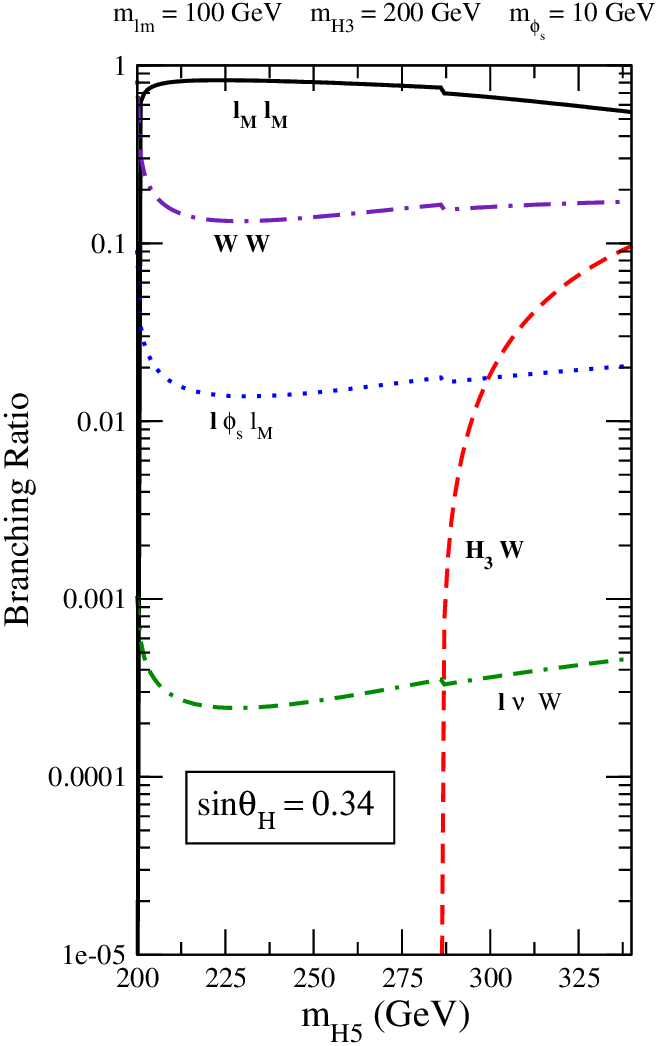}
    \includegraphics[width=5cm]{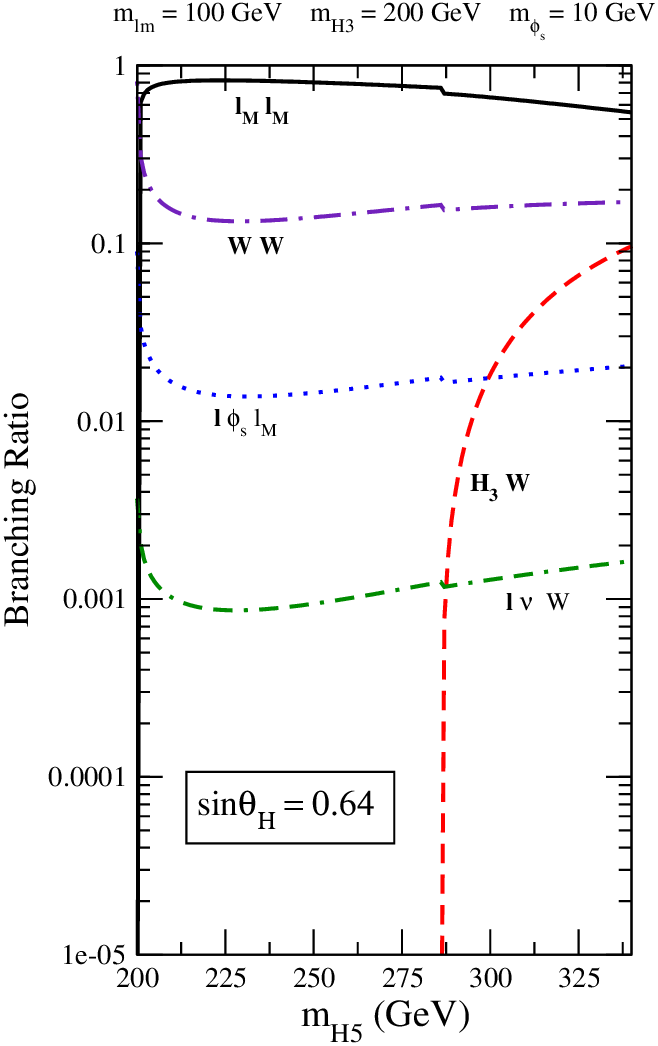}
    \includegraphics[width=5cm]{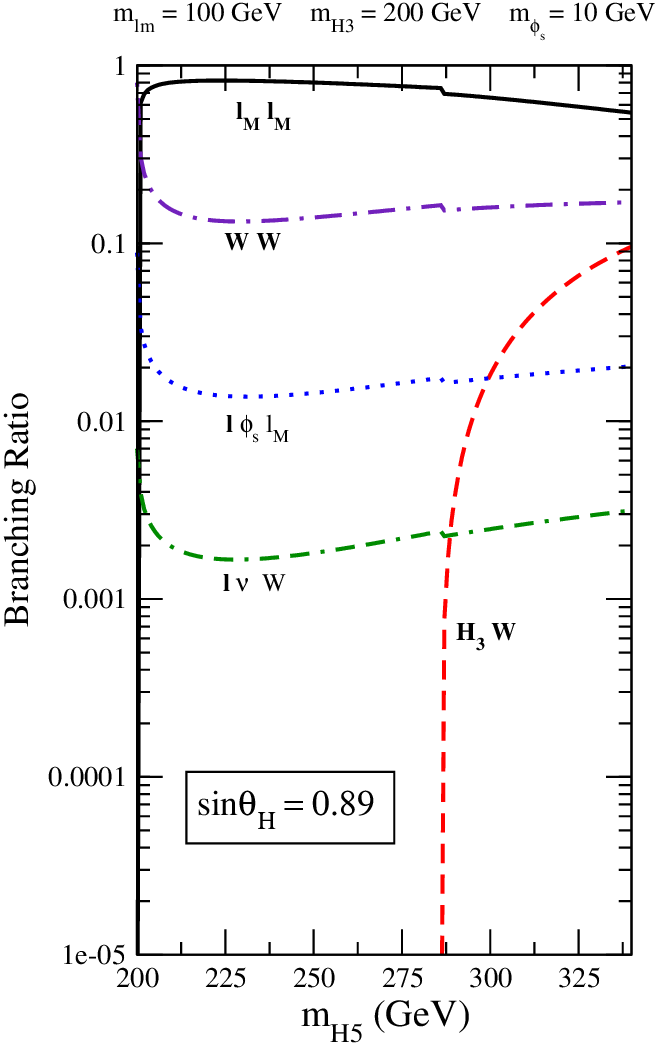}
    \caption{Same as before but with a heavier $m_{lM}$.}
    \label{fig:br2}
  \end{figure}
\end{center}

Similar results are obtained for larger $m_{lM}$ as can be seen in figure~\ref{fig:br2}
where we show the branching ratios for $m_{lM}=100\, {\rm GeV}$. 

\section{Conclusions}
We argue that the study of models with extended scalar sectors involving Higgs triplets
is well motivated. We study the phenomenology of a model that, using
Higgs triplets, can relate both EWSB and neutrino mass generation using an electroweak
scale seesaw mechanism. This can in principle make the seesaw mechanism testable at colliders.
The model offers a rich scalar phenomenology involving mirror fermions, a single scalar 
and the usual charged Higgs processes of extended Higgs models, in particular the doubly charged Higgs.
We have studied these processes in detail and have computed the branching ratios for the 
doubly charged Higgs. We find that for all the allowed parameter space, the dominant decay is
to the mirror fermions. The existence of this decay would provide a clean signature in 
favor of this scenario.

\acknowledgments
A.A. acknowledges support from CONACYT and SNI. A.A. and PQH also acknowledge the Aspen Center
for Physics for their hospitality while part of this work was being done.
J. H-S. was supported by CONACYT(M\'exico) under the grant J50027-F and by
PROMEP-grant (M\'exico).
PQH is supported in parts by the US Department of Energy under grant No.
DE-A505-89ER40518.

\newpage

\begin{thebibliography}{99}

\bibitem{Chanowitz:1985ug}
  M.~S.~Chanowitz and M.~Golden,
  %``Higgs Boson Triplets With M (W) = M (Z) Cos Theta Omega,''
  Phys.\ Lett.\  B {\bf 165}, 105 (1985).
  %%CITATION = PHLTA,B165,105;%%

\bibitem{georgi}
H.~Georgi and M.~Machacek,
Nucl.\ Phys.\ B {\bf 262}, 463 (1985).
%%CITATION = NUPHA,B262,463;%%


%\cite{Hung:2006ap}
\bibitem{Hung:2006ap}
  P.~Q.~Hung,
  %``A Model of electroweak-scale right-handed neutrino mass,''
  Phys.\ Lett.\  B {\bf 649}, 275 (2007)
  [arXiv:hep-ph/0612004].
  %%CITATION = PHLTA,B649,275;%%

%\cite{Accomando:2006ga}
\bibitem{Accomando:2006ga}
  E.~Accomando {\it et al.},
  %``Workshop on CP studies and non-standard Higgs physics,''
  arXiv:hep-ph/0608079.
  %%CITATION = HEP-PH/0608079;%%
  
  %\cite{Gunion:1989ci}
\bibitem{Gunion:1989ci}
  J.~F.~Gunion, R.~Vega and J.~Wudka,
  %``Higgs triplets in the standard model,''
  Phys.\ Rev.\  D {\bf 42}, 1673 (1990).
  %%CITATION = PHRVA,D42,1673;%%
  
  %\cite{Han:2007bk}
\bibitem{Han:2007bk}
  T.~Han, B.~Mukhopadhyaya, Z.~Si and K.~Wang,
  %``Pair Production of Doubly-Charged Scalars: Neutrino Mass Constraints and
  %Signals at the LHC,''
  arXiv:0706.0441 [hep-ph].
  %%CITATION = ARXIV:0706.0441;%%
  
\bibitem{Hung2}
  P.~Q.~Hung,
  %''Consequences of a Pati-Salam unification of the electroweak-scale active $\nu_R$ model,''
  Nucl. \ Phys. \ B {\bf 805}, 326 (2008) 
  arXiv:0805.3486[hep-ph].
  %%CITATION = ARXIV:0805.3486;%%
  
%\cite{Hung:2007ez}
\bibitem{Hung:2007ez}
  P.~Q.~Hung,
  %``Electroweak-scale mirror fermions, $\mu \to e \gamma$ and $\tau \to \mu
  %\gamma$,''
  Phys.\ Lett.\  B {\bf 659}, 585 (2008)
  [arXiv:0711.0733 [hep-ph]].
  %%CITATION = PHLTA,B659,585;%%

\bibitem{Haber}
  H.~E.~Haber and H.~E.~Logan,
  %``Radiative corrections to the Z b anti-b vertex and constraints on  extended
  %Higgs sectors,''
  Phys.\ Rev.\  D {\bf 62}, 015011 (2000)
  [arXiv:hep-ph/9909335].
  %%CITATION = PHRVA,D62,015011;%%
  
\bibitem{Aaki2007}
  M.~Aoki and S.~Kanemura,
  %``Unitarity bounds in the Higgs model including triplet fields with custodial
  %symmetry,''
  Phys.\ Rev.\  D {\bf 77}, 095009 (2008)
  [arXiv:0712.4053 [hep-ph]].
  %%CITATION = PHRVA,D77,095009;%%
  
\bibitem{LEP} 
  See Table 14 and discussion in S.~Schael {\it et al.}  
  [ALEPH Collaboration and DELPHI Collaboration and L3 Collaboration and ],
  %``Search for neutral MSSM Higgs bosons at LEP,''
  Eur.\ Phys.\ J.\  C {\bf 47}, 547 (2006)
  [arXiv:hep-ex/0602042].
  %%CITATION = EPHJA,C47,547;%%
  
  %\cite{Ellis:1991}
\bibitem{Ellis-Ridolfi}
  J.~R.~Ellis, G.~Ridolfi and F.~Zwirner,
  %``On radiative corrections to Supersymetric Higgs boson masses and their implications for
  %LEP searches,''
  Phys.\ Lett.\ B {\bf 257}, 83 (1991) and Phys.\ Lett.\ B {\bf 262}, 477 (1991).
  %  [hep-ph/9912516].
  %%CITATION = HEP-PH 9912516;%%
  
  %\cite{DiazCruz:2007tf}
\bibitem{DiazCruz:2007tf}
  J.~L.~Diaz-Cruz, J.~Hernandez-Sanchez, S.~Moretti and A.~Rosado,
  %``Charged Higgs boson phenomenology in Supersymmetric models with Higgs
  %triplets,''
  Phys.\ Rev.\  D {\bf 77}, 035007 (2008)
  [arXiv:0710.4169 [hep-ph]].
  %%CITATION = PHRVA,D77,035007;%%

\end{thebibliography}
\end{document}